\documentclass[aps,prd,superscriptaddress,eqsecnum,nofootinbib,showkeys,twocolumn,showpacs,preprintnumbers]{revtex4-1}
\usepackage{graphicx}
\usepackage{amsfonts}
\usepackage{amssymb}
\usepackage{amsmath}
\usepackage{flafter}
\usepackage{epstopdf}
\usepackage{hyperref}
\usepackage{xcolor}
\usepackage{float}
\usepackage[stable]{footmisc}

\begin{document}

\title{Laser Cosmology}

\date{\today} 
\author{Pisin Chen}
\affiliation{Department of Physics and Graduate Institute of Astrophysics, National Taiwan University, Taipei, Taiwan 10617}
\email{pisinchen@phys.ntu.edu.tw}
\affiliation{Leung Center for Cosmology and Particle Astrophysics (LeCosPA), National Taiwan University, Taipei, Taiwan, 10617}
\affiliation{Kavli Institute for Particle Astrophysics and Cosmology, SLAC National Accelerator Laboratory, Menlo Park, CA 94025, U.S.A.}

\addtocounter{footnote}{-20}

\begin{abstract}

Recent years have seen tremendous progress in our understanding of the cosmos, which
in turn points to even deeper questions to be further addressed. Concurrently the laser technology has undergone dramatic revolutions, providing exciting opportunity for science applications. History has shown that the symbiosis between direct observations and laboratory investigation is instrumental in the progress of astrophysics. We believe that this remains true in cosmology. Current frontier phenomena related to particle astrophysics and cosmology 
typically involve one or more of the following conditions: (1) extremely high energy events;(2) very high density, high temperature processes; (3) super strong field environments. Laboratory experiments using high intensity lasers can calibrate astrophysical observations, investigate underlying dynamics of astrophysical phenomena, and probe fundamental physics in extreme limits. In this article we give an overview of the exciting prospect of laser cosmology. In particular, we showcase its unique capability of investigating frontier cosmology issues such as cosmic accelerator and quantum gravity.

\vspace{3mm}


\end{abstract}

\maketitle

\section{Introduction}

This is an exciting time for astrophysics and cosmology.
New observations and results from space-based, ground-
based, and underground-based experiments are pouring
in by the day, which have greatly advanced our
knowledge of the universe, Modern particle astrophysics and cosmology lie at the intersection
between several fields of physics. Specifically, there
are fundamental issues that overlap cosmology with
particle physics, or that connect quarks with the cosmos (See Fig. 1). The present state of pursuit
in this frontier can perhaps be best summarized by the
ÒEleven Science Questions for the New CenturyÓ posted
by the U.S. National Research CouncilÕs Committee on
the Physics of the Universe \cite{Turner}. These are:

\noindent
1. What is the dark matter?

\noindent
2. What is the nature of the dark energy?

\noindent
3. How did the universe begin?

\noindent
4. Did Einstein have the last word on gravity?

\noindent
5. What are the masses of the neutrinos, and how
have they shaped the evolution of the universe?

\noindent
6. How do cosmic accelerators work and what are
they accelerating?

\noindent
7. Are protons unstable?

\noindent
8. Are there new states of matter at exceedingly
high density and temperature?

\noindent
9. Are there additional spacetime dimensions?

\noindent
10. How were the elements from iron to uranium
made?

\noindent
11. Is a new theory of matter and light needed at
the highest energies?

\begin{figure}
\includegraphics[width=8.5cm]{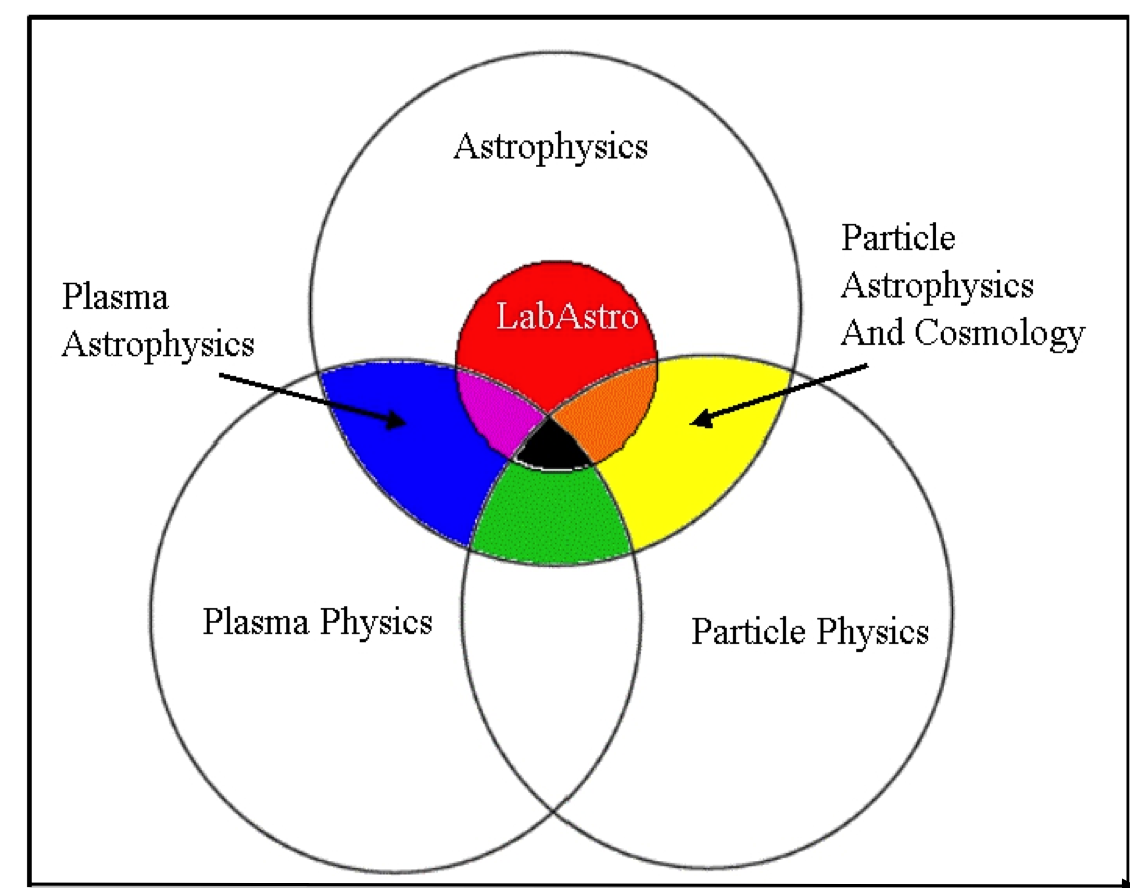}
\caption{A diagram that indicates the relationship between
laboratory astrophysics and astrophysics, particle physics and
plasma physics \cite{Chen2003}.}
\label{f1}
\end{figure}

There is also the astrophysical frontier that lies at
the intersection between astrophysics and plasma physics
(See also Fig. 1). It is known that the (ordinary) matter
in our universe largely exists in the plasma state. While
the study of plasma astrophysics has a long history, its
modern frontier typically involves very high density, high
pressure, and high temperature plasma processes. This
frontier lies in the domain of the newly emerged field of
high energy-density physics.
Astrophysical phenomena associated with the Eleven
Science Questions raised above often involve one or more
of the following extreme conditions \cite{Chen2003}:

\noindent$\bullet$
Extremely high energy events, such as ultra
high energy cosmic rays (UHECRs), neutrinos,
gamma rays, etc;

\noindent$\bullet$
Very high density, high pressure, high temper-
ature processes such as supernova explosions and
gamma ray bursts (GRBs);

\noindent$\bullet$
Super strong field environments, such as that
around black holes (BH) and neutron stars (NS).

Due to these connections, certain aspects of the particle
astrophysics issues are further linked with high energy-
density physics (See Fig. 1).

The history of laboratory astrophysics (LA) dates back for at least a century (for a brief historical recount, see \cite{Ciardi}). In recent years, there have been organized, programatic efforts to promote it. The importance of laboratory astrophysics was noted by the US National Research Council 2010 Decadal Survey of Astronomy and Astrophysics (Astro2010), which identified laboratory astrophysics as one of NASA's core research programs ``fundamental to mission development and essential for scientific progress". NASA organized a Laboratory Astrophysics Workshop in 2010 (LAW2010). The ``Laboratory Astrophysics White Paper" \cite{Savin} based on this workshop summarized the state of LA and its outlook. It primary scientific attentions are in astrophysics related to atomic, molecular, dust and ices, plasma phenomena. There have also been LA efforts along the line of ``high energy density physics" \cite{Drake2006}. Here the main attention is to the understanding of hydrodynamical systems such as jets and instabilities in astrophysical systems. Relativistic laser-plasma interactions is the means for such investigations. Such attempt has been strongly echoed by experts in the laser and plasma physics communities. For example, Esirkepov and Bulanov \cite{Esirkepov2012} advocate the powerfulness of ultra-intense lasers as a tool to investigate fundamental physics and relativistic laboratory astrophysics through laser-matter interaction. The studies include such frontier astrophysical dynamics as collisionless shocks, bow shocks, plasma wakes, magnetic reconnections, radiation pressure, etc. On the fundamental physics front, 
 the extreme power lasers can help investigate $e^+e^-\gamma$ plasma, laser-driven collider for quark-gluon plasma studies, etc. 

\section{Laser Intensity and its Probe of Matter}

The invention of chirped pulse amplification (CPA) scheme in the mid 1980s \cite{Strickland} has ushered in a second laser revolution since its inception on 1960 with dramatic increase in the laser peak power. It has become common place in present-day lasers to deliver a peak power beyond $10^{18} {\rm W/cm}^2$ at its focus, where electrons interacting with such a field can become relativistic within one laser field oscillation. State-of-the-art laser can reach a focused intensity at $10^{22} {\rm W/cm}^2$ \cite{Bahk2004}. The ELI project currently under construction is designed for producing femtosecond, 70 KJ pulses, corresponding to a peak power $> 100 {\rm PW}$ and an intensity $> 10^{25} {\rm W/cm}^2$. Figure 2 \cite{Esirkepov2012} shows the laser focused intensity as a function of the year and the corresponding reach of the state of matter.  We see that the state of matter that can be probed by lasers progresses from atoms, molecules, plasmas, to relativistic and quantum regimes of plasmas, and finally approaching the so-called Schwinger limit, at $10^{30} {\rm W/cm}^2$, where the QED vacuum becomes so unrest that spontaneous $e^+e^-$ pair production, so-called ``boiling of vacuum", should occur \cite{Chen1998}. The intensity of the state-of-the-art lasers, however, is still about 5 orders of magnitude below the Schwinger limit. 

\begin{figure}
\includegraphics[width=8.5cm]{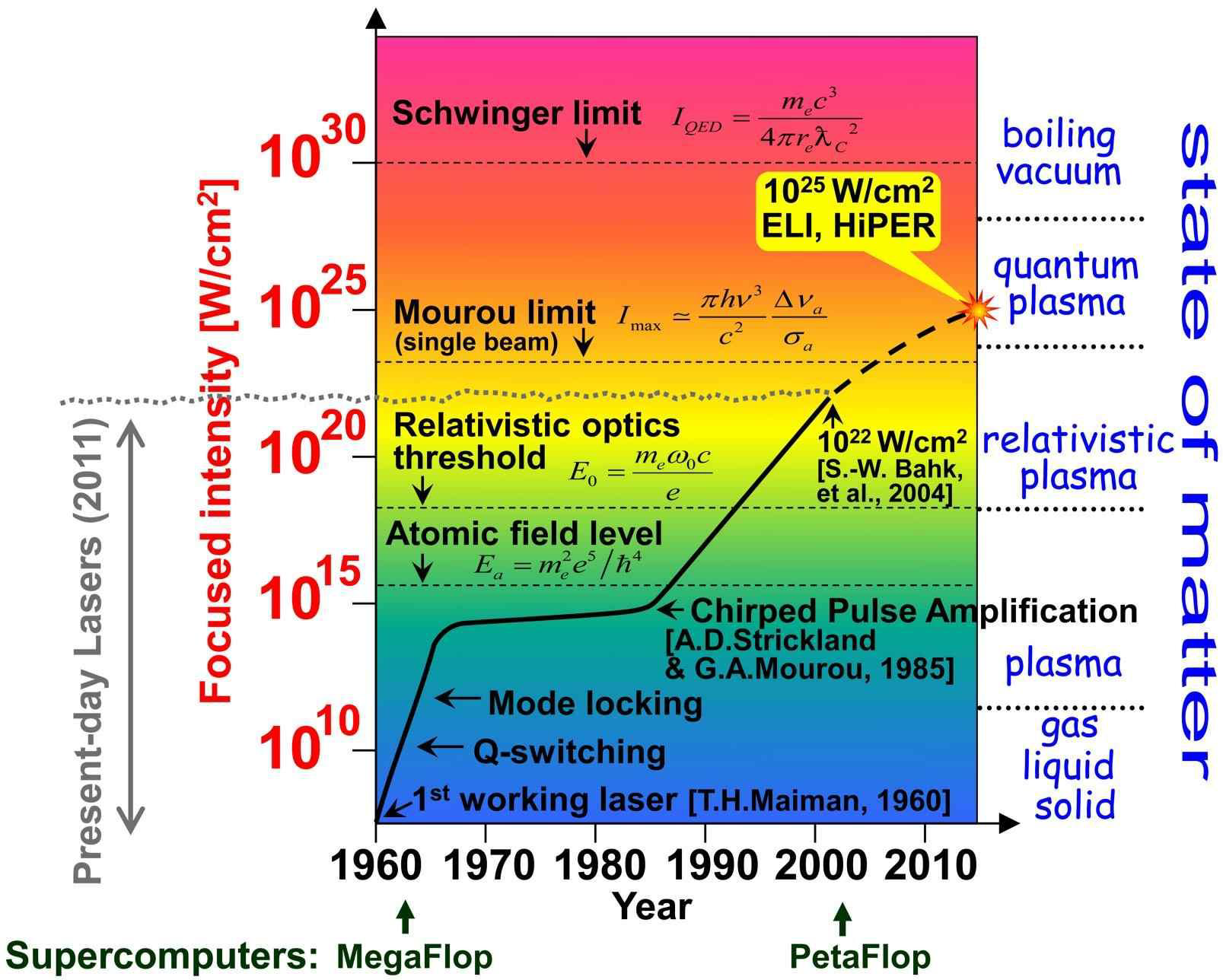}
\caption{Increase in laser intensity as a function of time. The state of matter that can be probed by certain laser intensity is indicated on the right. (Figure reproduced from \cite{Esirkepov2012})}
\label{f2}
\end{figure}

Most recently, there is yet another exciting breakthrough of laser technology, namely the {\it fiber accelerator} \cite{Mourou2013}. This new approach provides much hope for eventually reaching the Schwinger limit. At this limit, in addition to the probe of nonlinear QED effects including the boiling of the vacuum, perhaps the more exciting science potential is the probe of quantum gravity effects such as the celebrated black hole (BH) Hawking radiation \cite{Hawking} through its analog in flat space-time, the Unruh effect \cite{Unruh}, via violent acceleration driven by the ultra-intense lasers \cite{ChenTajima}. The use of ultra-intense lasers to probe quantum gravity effects opens up an exciting new prospect. This inspires us to introduce a new terminology, {\it laser cosmology}, to refer to the study of cosmology using the ultra-intense lasers.

\section{Three Aspects of Laser Cosmology}

In our view, there are three functional aspects of laser cosmology. Each has a different utility but all three are complementary to each other. We briefly introduce them in the following. 

\noindent
{\bf 1. Calibration of cosmic observations}

Traditionally, this aspect of laser cosmology has been the dominant content of laboratory astrophysics \cite{Savin}. Most astrophysical observations rely on photons induced from atomic or molecular spectra, synchrotron radiation, Compton scattering, fluorescence, electron-positron pair production, plasma oscillations, etc. Detailed calibrations of these in the laboratory is instrumental in validating the astrophysical observations. Novel observational techniques can also be qualified in the laboratory setting. 

Though mundane, the scientific value of this aspect of laser cosmology is most certain. 

\noindent
{\bf 2. Validation of astrophysical dynamics}

The value of this line of efforts lies in the revelation of the dynamical underpinnings in the cosmos. 

Most astrophysical conditions are difficult to recreate in the laboratory setting. Luckily many astrophysical processes involve plasmas or magneto-hydrodynamics (MHD). which are in general scalable. Good examples are hydrodynamic jets, turbulences, shocks, instabilities, etc \cite{Drake2006,Esirkepov2012}. Through the controlled experiments in laboratories and the interplay with computer simulations, the plasma or MHD dynamics can be extended to the astrophysical scales and outside the range of the lab setting. 

The use of intense lasers to investigate astrophysical plasma or MHD processes has been nicely reviewed in, e.g., Ref. 5. In addition to those areas, laser cosmology can in principe address one of the 11 science questions listed above,  i.e., ``How do cosmic accelerators work and what are they accelerating?" We will review this topic in the following section as a showcase of laser cosmology.

\noindent
{\bf 3. Investigation of fundamental cmosic physics}

In the previous two aspects, typically the underlying physical principles are part of the established knowledge in physics, while the specific details, due either to the complexity of the astrophysical environment or the largeness of the astrophysical scales, require validation in the laboratory. In contrast, this last aspect of laser cosmology often deals with physical problems whose foundations have not yet been established. More importantly, some of these issues, though extremely fundamental and essential to cosmology, are impossible to observe in the cosmos. 

One good example is the famous black hole Hawking evaporation, which serves as a crucial window to look into the nature of quantum gravity. Yet safe other than that for primordial BHs, the Hawking temperature for typical stellar size BH or supermassive BH at galactic center are too low and its radiation too faint to observe in the foreseeable future. While the laboratory investigation of quantum gravity effect  is indirect, it may be the best tool we have at hand to experimentally study this very important issue in quantum gravity, the understanding of which may shed lights on many other issues in cosmology. 

As the topics in laser cosmology cover a broad spectrum, it is quite impossible to review all aspects in a short article. We instead review two topics, each relates to one of the 11 science questions, as our highlights. 

\section{Cosmic Acceleration}
\noindent
{\it How do cosmic accelerators work and what are they accelerating?}

Conventional cosmic acceleration mechanisms are based on the (second order) Fermi acceleration \cite{Fermi} and its variant, the diffusive shock acceleration \cite{DSA}, which is a first-order Fermi mechanism. The former is based on the energy gain of a proton or an ion when it bounces off a magnetic field domain. The latter is based on multiple crossing of shock fronts by the particle in a relativistic astrophysical outflow. Both mechanisms face challenges when applied to the acceleration of ultra high energy cosmic ray (UHECR) particles. For the original Fermi mechanism, an ever-increasing magnetic field strength and domain is required, but it faces the limit. For the diffusive shock acceleration, the mechanism is collisional, that is the particle has to turn around (in the rest frame of the shock) and cross the shock front multiple times through the collision with, e.g., magnetic bubbles in the outflow medium. At extremely high energies, such as $10^{19} {\rm eV}$, even a proton would have a Lorentz factor $\gamma\sim 10^{10}$! Any bending of the trajectory of a proton at such energy would necessarily induce severe radiation loss. It is evident that new thinking is needed in terms of cosmic accelerator for UHECR.
 
Based on the above discussion, one guiding principle for UHECR accelerator is that it should be {\it linear} to avoid any unnecessary radiative energy loss. 
Two proposed new mechanisms, the Zevatron \cite{Blandford2001} and the plasma wakefield acceleration \cite{Chen2002}, belong to this linear acceleration category. The idea behind Zevatron is uni-polar induction acceleration. It was proposed that the AGN jet can carry with it electro-static fields along the axis and the `battery circuit' is closed via the return current along the equatorial plane. It was shown that this induction acceleration could provide the required energetics of UHECR. 

Plasma wakefield acceleration mechanisms driven by lasers \cite{Tajima1979} or by high-energy particle beams \cite{ Chen1985} have been confirmed and actively pursued in the past several decades. In 2002, Chen, Tajima and Takahashi \cite{Chen2002} applied it as the underlying mechanism for cosmic accelerator. It was suggested that plasma medium waves or shocks in an astrophysical relativistic outflow can also induce wakefileds to accelerate particles. The stochastic encounter of acceleration-deceleration phases would result in a power-law energy spectrum that scales as $E^{-2}$, just like the Fermi mechanism. In fact, it was shown that any purely stochastic acceleration process would follow this energy spectrum. The confirmation of this media-wave driven plasma wakefield generation and acceleration was demonstrated by Chang et al. \cite{Chang} and Hoshino et al. \cite{Hoshino}. More recently Ebisuzaki and Tajima proposed a variant acceleration of UHECR by the wakefield excited by the Alvfen waves propagating along the AGN jets, which can accelerate protons/nuclei to extreme energies beyond $10^{21}$ eV \cite{Ebisuzaki}.

\section{Quantum Gravity}
\noindent
{\it Did Einstein have the last word on gravity?}

Einstein's general theory of relativity (GR) has been one of the most successful theory in physics. Phenomena such as gravitational lensing, black holes, Hubble expansion, inflation, to late-time accelerating expansion of the universe can all be accommodated by GR. On the other hand, the universe is fundamentally quantum mechanical (QM). But so far a self-consistent quantum theory of gravity is still lacking. In the pursuit of a final theory of quantum gravity, the BH Hawking evaporation effect serves as a unique window to help unravel the deeper connections between GR and QM. Indeed, the temperature associated with the blackbody Hawking radiation, experienced by a stationary observer outside the BH, manages to connect gravity, QM and statistical mechanics in one stroke \cite{Hawking}:
\begin{equation}
k_BT_H=\frac{\hbar g}{2\pi c}=\frac{\hbar c^3}{8\pi G M},
\end{equation}
where $g$ is the gravitational acceleration at the BH event horizon and $M$ is the black hole mass (see Fig. 3). We note that $T_H$ is inversely proportional to the BH mass. Unfortunately this temperature is extremely low for stellar size and supermassive BHs. Primordial BHs borne out of space-time fluctuations, which are small, might have its Hawking effect detectable. However, this would only happened in the very early universe and therefore unobservable in the present day. Giving the unique window provided by the Hawking effect to peak into the cross road between GR, QM, and therefore mutual connections with thermodynamics, this is an opportunity that should not be easily given away.

It happens that a similar effect by the name of Unruh effect, may help to shed some light on the problem at stake. As demonstrated by Unruh \cite{Unruh}, a uniformly accelerated observer would see the surrounding vacuum develops a heat bath with a blackbody temperature of 
\begin{equation}
k_BT_U=\frac{\hbar a}{2\pi c},
\end{equation}
where $a$ is the observer's proper acceleration. Although the space-time involved in the Unruh effect is flat and therefore it is not the same as the Hawking effect, the physics behind the two effects are similar, namely the existence of the event horizon in both cases that causes the entanglement of entropy across the horizon that gives rise to a temperature that looks remarkably similar in the two cases (see Fig. 3). 

\begin{figure}
\includegraphics[width=8.5cm]{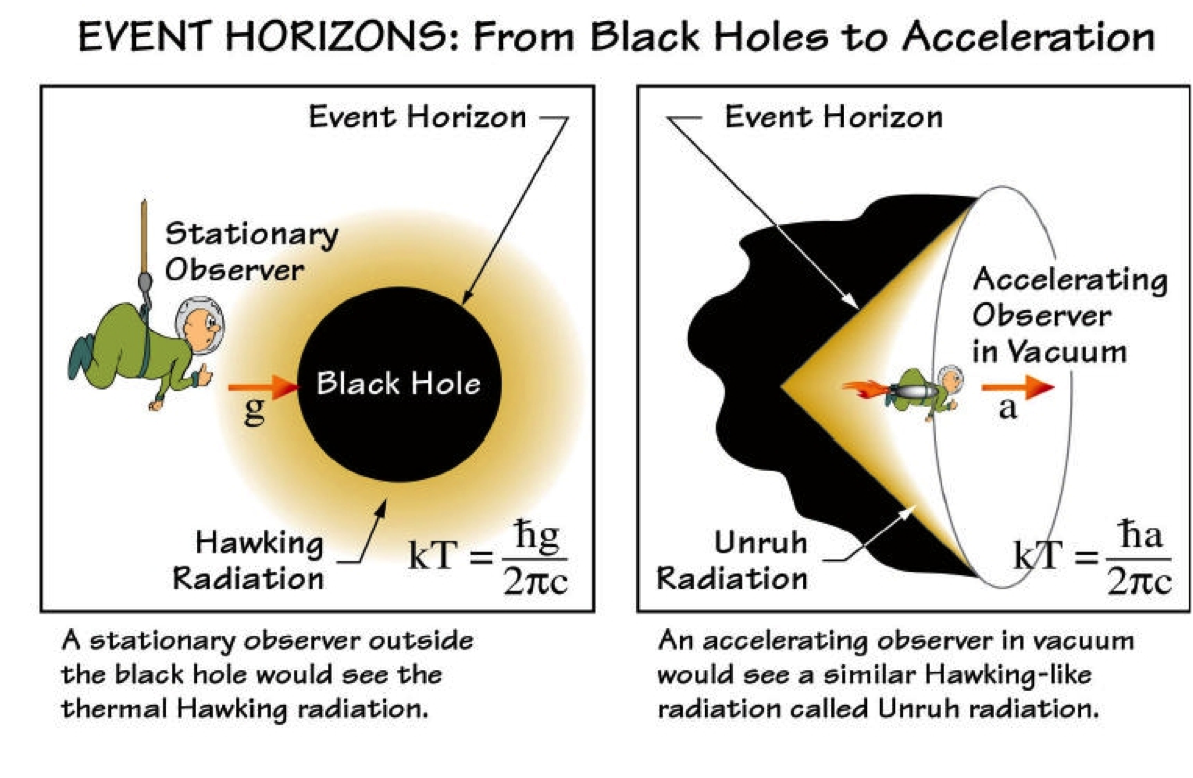}
\caption{A comparison between the black hole Hawking effect and a flat space-time analogy of Unruh effect}
\label{f3}
\end{figure}

The dramatic advancement of laser technology in recent years provides an opportunity to test the Unruh effect via the violent acceleration induced by the ultra-intense electromagnetic fields of the laser pulse. Motivated by this, Chen and Tajima \cite{ChenTajima} proposed a laser experiment on the Unruh effect. They suggested the setup of two identical, counter-propagating linearly polarized laser pulses, either in a standing-wave or transient configuration. In this setup there exist nodal points where the laser electric fields are maximum while the magnetic fields are minimum. A `particle detector', in tho case an electron, located at a nodal point would suggest to maximum acceleration with minimum bending. The accelerated electron would observe a thermal heat bath in its surrounding, and its interaction with the heat bath would trigger photon emissions, which is isotropic in the electron's rest frame. In the lab frame this isotropic radiation would be forward-boosted that peaks along the direction of acceleration with a half width angle $\sim 1/\gamma$. In contrast, the competing Larmor radiation background, which is a dipole radiation in the electron's rest frame and in a cone shape in the lab frame with a cone angle $\sim 1/\gamma$. This means there exists a `blind spot' of the Larmor radiation in the lab frame where the Unruh signals happen to be the strongest. This suggests that the testing of Unruh effect using ultra-intense lasers maybe conceivable (see Fig. 4).   

\begin{figure}
\includegraphics[width=8.5cm]{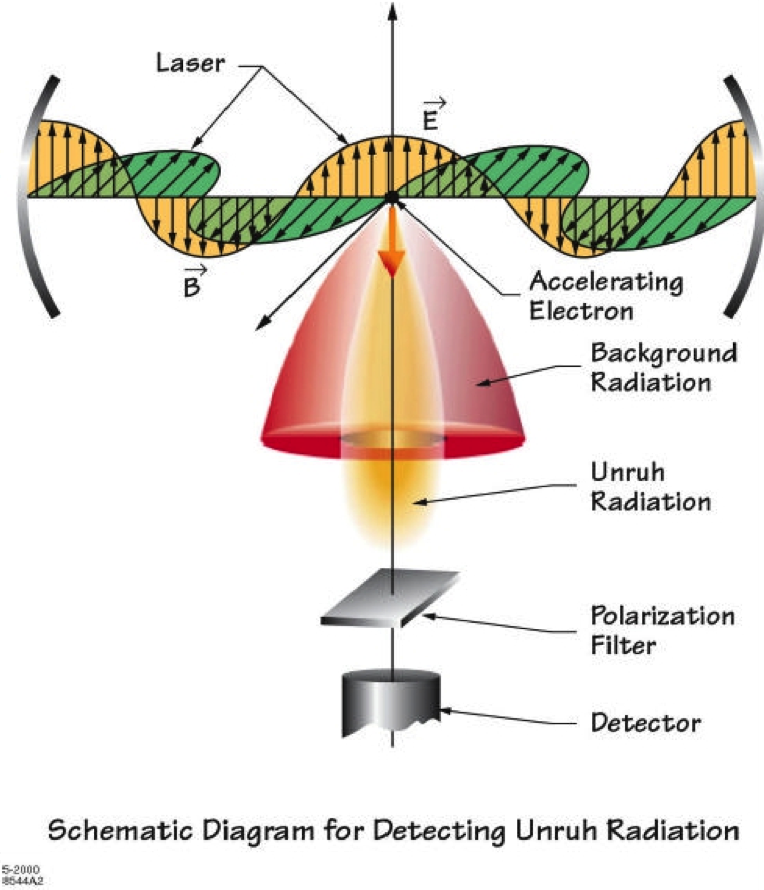}
\caption{A schematic diagram for a proposed laser test of Unruh effect based on laser standing wave configuration.}
\label{f4}
\end{figure}

Additional considerations along this line of thought indicates that the emitted Unruh photons are created in pairs whose polarizations are perfectly correlated \cite{Schutzhold}. Recently, a more realistic system based on sinusoidal time-dependent of the laser electric field was considered \cite{Doukas2013}. It was pointed out that in principle the laser-induced acceleration is nonuniform and thus the horizon does not exist, unlike the situation for the bone fide Rindler geometry. Nevertheless the accelerated particle detector still sees a nontrivial effective vacuum Unruh temperature. The time-averaged proper acceleration for the case of standing wave oscillatory electric field reads
\begin{equation}
\bar{a}=\frac{\omega \sinh^{-1}2a_0}{F(\pi/2,-4a_0^2)},
\end{equation}
where $\omega$ is the laser frequency, $a_0\equiv eE_0/m\omega$ the laser strength parameter and $F(\phi,m)$ is the elliptic integral of the first kind. In the low frequency limit and at high time-averaged proper acceleration, the effective temperature approaches the Unruh temperature:
\begin{equation}
k_B\bar{T}_U=\frac{\hbar \bar{a}}{2\pi c},
\end{equation}
whereas at high frequencies the effective temperature in the laser experiment is somewhat lower than the bone fide Unruh temperature. 

We see that since the original proposal of a laser test of Unruh effect, the concept has been gradually refined. On the whole, the effective Unruh temperature is evidently in favor of high accelerations, and that corresponds to the development of highest possible laser powers. 

\section{Summary}

In this article we have briefly overviewed the 11 fundamental questions facing frontier cosmology. We then reviewed the different roles of laboratory astrophysics and pointed out that some of these challenges maybe addressed by laser cosmology. With the dramatic increase of laser peak power in recent years, the time is ripe to use ultra-intense lasers to investigate astrophysics and cosmology under extreme conditions. The most exciting prospect, however, is the possibility of probe quantum gravity with intense lasers. As argued, critical issues such as the black hole Hawking effect cannot in principle be observed directly from sky, which renders the role of laser cosmology even more persuasive. 

\vskip 0.5cm
\noindent{\bf Acknowledgement}

This work is supported by Taiwan's National Science Council and the Leung Center for Cosmology and Particle Astrophysics (LeCosPA) of National Taiwan University.

\end{document}